\documentstyle[namedreferences,epsfig]{spackap}

\input epsf

\newcommand{\gsim}{\gas}
\newcommand{\lsim}{\las}

\def\beq{\begin{equation}}
\def\eeq{\end{equation}}

\def\alwaysmath#1{{\ifmmode{#1}\else{$#1$}\fi}}

\def\hii{H\thinspace{$\scriptstyle{\rm II}$}}
\def\etal{{\it et al.}}

\begin{opening} 
\title{NON-BBN CONSTRAINTS ON THE KEY COSMOLOGICAL PARAMETERS} 
 
\author{Gary \surname{Steigman}}         
\institute{Departments of Physics and Astronomy \\ The Ohio State University, 
Columbus, OH 43210, USA}
\author{Naoya \surname{Hata}}         
\institute{Institute for Advanced Study, Princeton, NJ 08540}
\author{James E. \surname{Felten}}
\institute{Code 685, NASA Goddard Space Flight Center, Greenbelt, MD 20771} 
 
\date{}

\end{opening} 

\begin{ao}
Professor G. Steigman, Department of Physics, The Ohio State University, 
174 West 18th Avenue, Columbus, OH 43210, USA 
\end{ao}                   

\runningauthor{G. Steigman, N. Hata, and J. E. Felten}
\runningtitle{NON-BBN CONSTRAINTS ON THE KEY COSMOLOGICAL PARAMETERS}

\begin{document}

\begin{abstract} 
Since the baryon-to-photon ratio $\eta_{10}$ is in some 
doubt at present, we ignore the constraints on $\eta_{10}$ 
from big bang nucleosynthesis (BBN) and fit the three key 
cosmological parameters ($h, \Omega_{\rm{M}}, \eta_{10}$) 
to four other observational constraints: Hubble parameter 
($h_{\rm{o}}$), age of the universe ($t_{\rm{o}}$), cluster 
gas (baryon) fraction ($f_{\rm{o}} \equiv f_{\rm{G}}h^{3/2}$), 
and effective shape parameter ($\Gamma_{\rm{o}}$).  We 
consider open and flat CDM models and flat $\Lambda$CDM models, 
testing goodness of fit and drawing confidence regions by the 
$\Delta\chi^2$ method.  CDM models with $\Omega_{\rm{M}} =1$ 
(SCDM models) are accepted only because we allow a large error 
on $h_{\rm{o}}$, permitting $h < 0.5$.  Open CDM models are 
accepted only for $\Omega_{\rm{M}} \gsim 0.4$.  $\Lambda$CDM 
models give similar results.  In all of these models, large 
$\eta_{10}$($\gsim 6$) is favored strongly over small 
$\eta_{10}$($\lsim 2$), supporting reports of low deuterium 
abundances on some QSO lines of sight, and suggesting that 
observational determinations of primordial $^4$He may be 
contaminated by systematic errors.  Only if we drop the 
crucial $\Gamma_{\rm{o}}$ constraint are much lower 
values of $\Omega_{\rm{M}}$ and $\eta_{10}$ permitted.

\end{abstract} 


\section{Introduction} 

In the context of the hot big bang cosmology, if the 
number of light-neutrino species has its standard value 
$N_\nu =3$, the predicted primordial abundances of four 
light nuclides (D, $^3$He, $^4$He, and $^7$Li) depend on
only one free parameter, $\eta_{10}$, the universal ratio
(at present) of nucleons (baryons) to photons (in units 
$10^{-10}$).  In principle, $\eta_{10}$ is overdetermined
by the observed or inferred primordial abundances of the 
four light nuclides.  Indeed, Steigman, Schramm, and Gunn
(1977) have exploited this fact to use big bang nucleosynthesis 
(BBN) to constrain
$N_\nu$.  The status quo ante is that observations,
principally of D and $^4$He, have rendered $\eta_{10}$ 
one of the best known of the key cosmological parameters:
$\eta_{10}$ = $3.4 \pm 0.3$ (Walker \etal, 1991; the error 
bars being roughly ``1$\sigma$'').  

At present, when the microwave background temperature 
$T$ = 2.728 K (Fixsen \etal, 1996), the universal baryonic 
mass-density parameter $\Omega_{\rm{B}}\; (\,\equiv 8 \pi G \rho_{\rm{B}}/3H_0^2\,)$ is related to $\eta_{10}$ by
\begin{equation} 
\Omega_{\rm{B}}\,h^2 
= 3.667 \times 10^{-3} \; \eta_{10} = 0.0125 \pm 0.0011. 
\label{Eq:Omega_B} 
\end{equation} 
However, recently there have emerged reasons to suspect that 
$\eta_{10}$ may not be so well determined, and even that the 
standard theory of BBN may not provide 
a very good fit to the current data (Hata \etal, 1995).  There 
are several options available for resolving this apparent conflict 
between theory and observation.  Although some change in standard 
physics could offer resolution (e.g., a reduction in the effective 
value of $N_\nu$ during BBN below its standard value 3; cf. Hata 
\etal, 1995), Hata \etal\ (1995) note that large systematic errors 
may compromise the abundance data (cf. Copi, Schramm, and Turner, 
1995). 
 
This controversy has been sharpened by new observations giving 
the deuterium abundances on various lines of sight to high-redshift 
QSOs. In principle, these data should yield the primordial D 
abundance, but current results span an order of magnitude.  If 
the low value (D/H by number $\approx 2 \times 10^{-5}$; Tytler, 
Fan, and Burles, 1996; Burles and Tytler, 1996) is correct, then 
$\eta_{10} \approx 7$ in the standard model, but then it seems 
impossible to reconcile the inferred abundance of $^4$He [Olive 
and Steigman, 1995 (OS)] with (standard) BBN for this value of 
$\eta_{10}$ unless there are large systematic errors in the 
$^4$He data.  If, instead, the high figures (D/H $\approx 2 \times 
10^{-4}$; Carswell \etal,  1994; Songaila \etal, 1994; Rugers and 
Hogan, 1996) are correct, then D and $^4$He are consistent with 
$\eta_{10} \approx 2$, but modellers of Galactic chemical evolution 
have a major puzzle: How has the Galaxy reduced D from its high 
primordial value to its present (local) low value without producing 
too much $^3$He (Steigman and Tosi, 1995), without using up too 
much interstellar gas (Edmunds, 1994; Prantzos, 1996), and without 
overproducing heavy elements (cf. Tosi, 1996, and references therein)?  
It appears that $\eta_{10}$, though known to order of magnitude, 
may now be among the less well-known cosmological parameters.  
Despite this, large modern simulations which explore other 
cosmological parameters are often limited to a single value 
of $\eta_{10} = 3.4$ (e.g., Borgani \etal, 1997). 
 
Given this unsettled situation Steigman, Hata, and Felten 
(1997; hereafter SHF) have proposed that it may be constructive 
to abandon nucleosynthetic constraints on $\eta_{10}$ entirely 
and to put $\eta_{10}$ onto the same footing as the other 
cosmological free parameters, applying joint constraints on 
all these parameters based on other (non-BBN) astronomical 
observations and on theory and simulation.  Armed with $\eta_{10}$ 
determined in this manner we may then ``predict" the primordial 
abundances of the light nuclides and compare with the data to 
test the consistency of standard BBN.  In this contribution to 
the proceedings of the ISSI Workshop on Primordial Nuclei and 
Their Galactic Evolution (Bern, Switzerland, 6--10 May 1997)
we present a brief description of our approach along with a 
summary of our results for a ``standard" (fiducial) choice of 
the observational constraints. For further details (especially 
of the many variations on the standard case to be described 
herein) and references the reader is encouraged to consult SHF.

\section{The Method: An Overview}

Our approach is to let the three key cosmological 
parameters ($h, \Omega_{\rm{M}}, \eta_{10}$) range 
freely, fit the constraints (observables other than 
nucleosynthetic) to be described below, test goodness 
of fit by $\chi^2$, and draw formal confidence regions 
for the parameters by the usual $\Delta\chi^2$ method.   
Most of the SHF results are not surprising, and related 
work has been done before (White \etal, 1996; Lineweaver 
\etal, 1997; White and Silk, 1996; Bludman, 1997), but not 
with these three free variables and the full $\chi^2$ 
formalism.  Further, some recent cosmological observations 
and simulations, particularly those related to the ``shape 
parameter'' $\Gamma$ and the cluster baryon fraction (CBF), 
seem to pose a challenge to popular models, and there is 
some doubt whether any simple model presently fits all data 
well.  Our approach, which begins by discarding nucleosynthetic 
constraints, provides a new way of looking at these problems.  
For example, the CBF and $\Gamma$ constraints have not been 
applied jointly in earlier work which often also adopts a 
precise value for $\eta_{10}$ (or $\Omega_{\rm{B}}$).  

SHF find that, given their conservative (generous) choice of 
error bar on $h$, the SCDM model is disfavored but by no means 
excluded.  But even with this generous error bar, large values 
$(\gsim 6)$ of $\eta_{10}$ ($\Omega_{\rm{B}}\,h^2 \gsim 0.022$) 
are favored strongly over low values ($\lsim 2$; $\Omega_{\rm{B}}
\,h^2$ $\lsim 0.007$).  This suggests that the low D abundances 
measured by Tytler \etal\ (1996) and by Burles and Tytler (1996) 
may be correct, and that the observed (extrapolated) primordial 
helium-4 mass fraction [$Y_{\rm{P}} \approx 0.23$; cf. OS and 
Olive, Skillman, and Steigman, 1997 (OSS)], thought to be well 
determined, may be systematically too low for unknown reasons. 

\section{CDM Models:  Parameters and Observables} 

\subsection{Parameters} 
 
The CDM models we consider are defined by three free parameters: 
Hubble parameter $h$; mass-density parameter $\Omega_{\rm{M}} 
= 8 \pi G \rho_{\rm{M}}/3H_0^2$; and baryon-to-photon ratio 
$\eta_{10}$, related to $\Omega_{\rm{B}}$ by equation~(\ref{Eq:Omega_B}).  
Here $\Omega_{\rm{M}}$ by definition includes all ``dynamical mass'': 
mass which behaves dynamically like ordinary matter in the universal 
expansion; $\Omega_{\rm{M}}$ is not limited to clustered mass only.  
Other free parameters having to do with structure formation, such 
as the tilt parameter $n$, could be added (White \etal, 1996; 
Kolatt and Dekel, 1997; White and Silk, 1996), but generally we 
have tried to avoid introducing many free parameters. 
 
\subsection{Observables} 
 
We consider four observables (constraints) with measured values 
and with errors which are assumed to be Gaussian: (1) Hubble 
parameter $h_{\rm{o}}$; (2) age of the universe $t_{\rm{o}}$; 
(3) gas-mass fraction $f_{\rm{o}} \equiv f_{\rm{G}} h^{3/2}$ 
in rich clusters; and (4) ``shape parameter'' $\Gamma_{\rm{o}}$ 
from structure studies.  In SHF we considered a fifth constraint: 
the dynamical mass-density parameter $\Omega_{\rm{o}}$ as inferred 
from cluster measurements or from large-scale flows.  Here, we 
ignore this constraint but we shall comment on its relation to 
our ``standard" results.

\subsubsection{Observed Hubble Parameter $h_{\rm{o}}$} 
 
For the Hubble parameter the observable $h_{\rm{o}}$ is simply 
fit with the parameter $h$.  Measurements of $h$ still show 
scatter which is large compared with their formal error estimates 
(Bureau, Mould, and Staveley-Smith, 1996; Tonry \etal, 1997; Kundi\'{c} 
\etal, 1997; Tammann and Federspiel, 1997).  This indicates systematic 
errors.  To be conservative (permissive), we adopt $h_{\rm{o}} = 
0.70 \pm 0.15$.  Perhaps a smaller error could be justified; below 
we will comment on the consequences of shrinking the error bar. 
 
\subsubsection{Observed Age of the Universe $t_{\rm{o}}$} 
 
The age for these $\Lambda = 0$ models is a function of $h$ and
$\Omega_{\rm{M}}$ given by: $t = 9.78\:h^{-1}\times f(\Omega_{\rm{M}};
\Lambda$=0){ Gyr} [Weinberg, 1972, equations (15.3.11) \& (15.3.20)].  
We take the observed age of the oldest globular clusters as 
$t_{\rm{GC}} = 14 \pm 2$ Gyr (Bolte and Hogan, 1995; Jimenez, 1997; 
D'Antona, Caloi, and Mazzitelli, 1997; Chaboyer \etal, 1997; Cowan 
\etal, 1997; cf. Nittler and Cowsik, 1997).  The universe is older 
than the oldest globular clusters by an unknown amount $\Delta t$.  
Although most theorists believe that $\Delta t$ must be quite small 
(1 or 2 Gyr at most), we are unaware of any conclusive argument 
which guarantees this.  To keep things simple, SHF introduced 
asymmetric error bars: $t_{\rm{o}} = 14^{+7}_{-2}$ Gyr, allowing 
enough extra parameter space at large ages to accommodate a 
conservative range of $\Delta t$; extremely large ages will be 
eliminated by the $h_{\rm{o}}$ constraint in any case.

\subsubsection{Observed Gas Mass Fraction in Clusters $f_{\rm{o}}$} 

As is suggested by simulations, we assume that rich clusters provide
a fairly unbiased sample of {\bf the universal ratio of baryonic to
dark matter}.  Thus, we use the cluster (hot) gas fraction, $f_{\rm{G}}$,
not as a constraint on $\Omega_{\rm{M}}$, but as a constraint on the 
universal baryon fraction, the ratio $\Omega_{\rm{B}}/\Omega_{\rm{M}}$.
We emphasize that the following argument assumes that rich clusters 
provide a fair sample of the universal baryon fraction but does 
{\bf not} assume that most of the mass in the universe, or any 
specific fraction of it, is in rich clusters.

The measurement of $f_{\rm{G}}$ poses some problems; for discussion
and references, see SHF.  We have followed the approach of Evrard, 
Metzler, and Navarro (1996), who used gas-dynamical simulations to 
model the observations.  They find that the largest contribution to
the error in $f_{\rm{G}}$ arises from the measurement of the 
cluster's {\it total} mass, and they suggest that this error can 
be reduced by using an improved estimator and by restricting the 
measurement to regions of fairly high overdensity.  Evrard (1997) 
applies these methods to data for real clusters and finds 
$f_{\rm{G}}\,h^{3/2} = 0.060 \pm 0.003$.  To be conservative, 
we double his error bars and adopt for our constraint
\begin{equation} 
 f_{\rm{o}} \equiv f_{\rm{G}}\,h^{3/2} = 0.060 \pm 0.006. 
\end{equation} 

We are interested in the baryonic mass fraction, $\Omega_{\rm{B}}/
\Omega_{\rm{M}}$, but even in rich clusters not all baryons are in 
the form of gas, and selection factors may operate in bringing baryons 
and dark matter into clusters.  White \etal\ (1993) introduced a 
``baryon enhancement factor'' $\Upsilon$ to describe these effects.  
$\Upsilon$ may be defined by 
\begin{equation} 
f_{G0} = \Upsilon \, \Omega_{\rm{G}}/\Omega_{\rm{M}}, 
\label{Eq:f_G0} 
\end{equation} 
where $\Omega_{\rm{G}}$ is the initial contribution of {\it gas} 
to $\Omega_{\rm{M}}$ (note that $\Omega_{\rm{G}} \le 
\Omega_{\rm{B}}$) and $f_{G0}$ is the gas mass fraction in the 
cluster immediately after formation.  $\Upsilon$ is really the 
{\it gas} enhancement factor, because the simulations do not 
distinguish between baryonic condensed objects if any (galaxies, 
stars, machos) and non-baryonic dark-matter particles.  All of 
these are lumped together in the term $(\Omega_{\rm{M}} - 
\Omega_{\rm{G}})$ and interact only by gravitation. 

If all the baryons start out as gas $(\Omega_{\rm{G}} = 
\Omega_{\rm{B}})$, and if gas turns into condensed objects 
only {\it after} cluster formation, then equation~(\ref{Eq:f_G0}) 
may be rewritten: 
\begin{equation} 
f_{\rm{G}} + f_{\rm{GAL}} =  
\Upsilon \,\Omega_{\rm{B}} / \Omega_{\rm{M}}, 
\label{Eq:f_G+f_GAL} 
\end{equation} 
where $f_{\rm{G}}$ is the present cluster gas-mass fraction and 
$f_{\rm{GAL}}$ the present cluster mass fraction in baryonic 
condensed objects of all kinds (galaxies, stars, machos).  White 
\etal\ (1993) took some pains to estimate the ratio 
$f_{\rm{G}}/f_{\rm{GAL}}$ within the Abell radius of the Coma 
cluster, counting only galaxies (no stars or machos) in 
$f_{\rm{GAL}}$.  They obtained 
\begin{equation} 
f_{\rm{G}}/f_{\rm{GAL}} = 5.5 \, h^{-3/2}. 
\label{Eq:f_G/f_GAL} 
\end{equation} 
This is large, so unless systematic errors in this estimate are very 
large, the baryonic content of this cluster (at least) is dominated by 
the hot gas.  Carrying $f_{\rm{GAL}}$ along as an indication of 
the size of the mean correction for all clusters, and solving 
equations~(\ref{Eq:f_G+f_GAL}) and (\ref{Eq:f_G/f_GAL}) for 
$f_{\rm{G}} h^{3/2}$, we find 
\begin{equation} 
f_{\rm{G}}  h^{3/2}  = [\,1 + (h^{3/2}/5.5)\,]^{-1} (\Upsilon \, 
\Omega_{\rm{B}}/\Omega_{\rm{M}})\,h^{3/2}, 
\label{Eq:f_Gh} 
\end{equation} 
where $\Omega_{\rm{B}}$ is given from $\eta_{10}$ and $h$ by 
equation~(\ref{Eq:Omega_B}). 

This is the appropriate theoretical function of the free parameters 
to fit to the observations.   We set $\Upsilon = 0.9$ in our 
``standard" case.  Although this is representative of results from 
simulations, Cen (1997) finds that the determination of $f_{\rm{G}}$ 
from X-ray observations may be biased toward high $f_{\rm{G}}$ by 
large-scale projection effects; i.e., the calculated $f_{\rm{G}}$ 
exceeds the true $f_{\rm{G}}$ present in a cluster by a bias factor 
which can be as large as 1.4.  Although Evrard \etal\ (1996) and 
Evrard (1997) have not observed such a bias in their simulations, 
SHF explored its effect on our analysis by using for $\Upsilon$, 
instead of 0.9, an ``effective value'' $\Upsilon \approx 0.9 \times 
1.4 \approx 1.3$. 
  
\subsubsection{Shape Parameter $\Gamma_{\rm{o}}$ from Large-Scale Structure} 

The last observable we use is the ``shape parameter'' $\Gamma$, which 
describes the transfer function relating the initial perturbation 
spectrum $P_{\rm{I}} (k) \propto k^n$ to the present spectrum 
$P(k)$ of large-scale power fluctuations, as observed, e.g., in 
the galaxy correlation function.  When the spectral index $n$ of 
$P_{\rm{I}} (k)$ has been chosen, $\Gamma$ is determined by 
fitting the observed $P(k)$.
 
Results of observations may be cast in terms of an ``effective shape 
parameter'' $\Gamma$ (White \etal, 1996) which we take as our 
observable.  Studies show that for the usual range of CDM models, with 
or without $\Lambda$, the expression for $\Gamma$ is  
\begin{equation} 
  \Gamma \approx \Omega_{\rm{M}} h \; \exp \left [ -  
  \Omega_{\rm{B}} - (h/ 0.5)^{1/2} (\Omega_{\rm{B}} / 
  \Omega_{\rm{M}} ) \right ]  - 0.32 \ (n^{-1}-1) 
\label{Eq:Gamma_th} 
\end{equation} 
(Peacock and Dodds, 1994; Sugiyama, 1995; Liddle \etal, 1996a,b; White 
\etal, 1996; Liddle and Viana, 1996; Peacock, 1997).  For $n \approx 1$, 
if $\Omega_{\rm{B}}$ and $\Omega_{\rm{B}} /\Omega_{\rm{M}}$ are small, 
we have $\Gamma \approx \Omega_{\rm{M}} h$.  The Harrison-Zeldovich 
(scale-invariant, untilted) case is $n = 1$, which we adopt for our 
standard case.  

In contrast to the ``standard" case in SHF, here we adopt the more
conservative choice (larger error bars) for the observed value of $\Gamma$,
\begin{equation} 
\Gamma_{\rm{o}} = 0.25 \pm 0.05 
\label{Eq:Gamma_obs} 
\end{equation} 
(cf. Peacock and Dodds, 1994; Maddox, Efstathiou, and Sutherland, 1996). 
This is based on the galaxy correlation function, and it assumes that 
light traces mass. Equations~(\ref{Eq:Gamma_th}) and (\ref{Eq:Gamma_obs}) 
imply, very roughly, that $\Omega_{\rm{M}} h \approx 0.25$. 

The shape-parameter constraint is in a sense the least robust of 
the constraints we have discussed since it is not part of the basic 
Friedmann model.  Rather, it depends on a theory for the primordial 
fluctuations and how they evolve.  If the Friedmann cosmology were 
threatened by this constraint, we believe that those who model 
large-scale structure would find a way to discard it.  Therefore 
SHF have also explored the consequences of removing this constraint
and replacing it with one on $\Omega_{\rm{M}}$ derived from the $M/L$ 
ratio in clusters and the luminosity density of the universe (Carlberg, 
Yee, and Ellingson, 1997) or one based on studies of large-scale flows 
around voids (Dekel and Rees, 1994; cf. Dekel, 1997).

\section{CDM Models:  Results} 

\subsection{Standard Constraints} 

For our standard case we have four observational constraints:
$h_{\rm{o}} = 0.70 \pm 0.15$, $ t_{\rm{o}} = 14^{+7}_{-2}$ 
Gyr, $f_{\rm{o}} \equiv f_{\rm{G}} h^{3/2} = 0.060 \pm 0.006$, 
and $\Gamma_{\rm{o}} = 0.25 \pm 0.05$.   For this standard case 
we assume $n = 1$ and $\Upsilon = 0.9$.  The results for our 
three cosmological parameters are displayed in Figures 1 and 2
where the 68\% and 95\% confidence regions (``CRs") are shown.
Also shown are the projected CRs obtained by computing $\chi^2$ 
for single observables alone, or for pairs of observables.  
These are not true CRs but are intended to guide the reader 
in understanding how the various constraints influence the 
closed contours which show our quantitative results.

\begin{figure}
\centerline{\psfig{file=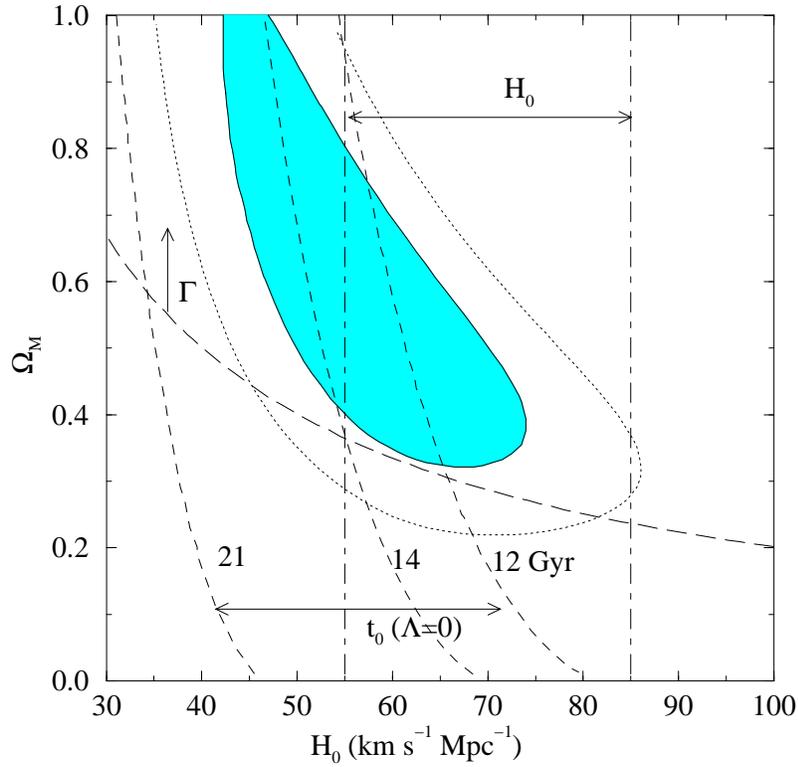,width=0.9\textwidth}}
\caption{68\% (shaded) and 95\% (dotted) confidence regions 
(``CRs'') in the $(H_0, \Omega_{\rm{M}})$ plane for CDM models 
with our four standard constraints.  The CRs are closed curves.  
Individual constraints in this plane are also shown schematically.}
\end{figure}

\begin{figure}
\centerline{\psfig{file=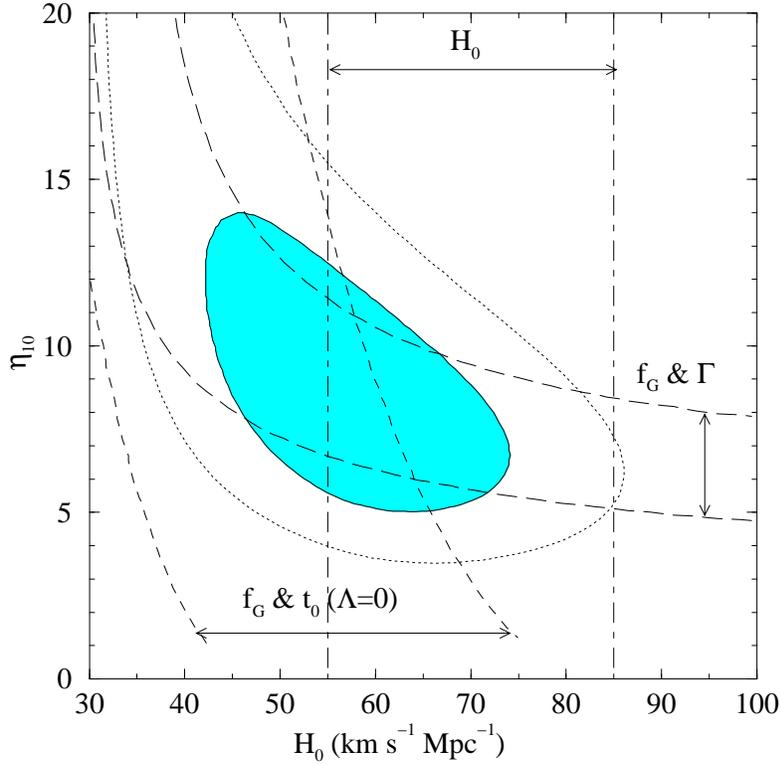,width=0.9\textwidth}}
\caption{Same as Figure 1, but in the $(H_0, \eta_{10})$ plane.  
Individual and paired constraints are also shown schematically.}
\end{figure}

Our best-fit values for the three key cosmological parameters are:
$\eta_{10} = 8.2^{+3.2}_{-2.2}$, $\Omega_{\rm{M}} = 0.48^{+0.22}_{-0.15}$ 
and $h = 0.58 \pm 0.22$.  Although the condition $\Omega_{\rm{M}} h 
\approx 0.25$ poses some threat to the SCDM $(\Omega_{\rm{M}} = 1)$ 
model, Figure 1 shows that this threat is far from acute given our more 
accurate form of the $\Gamma$ constraint in equation~(\ref{Eq:Gamma_th}), 
as long as the error on $h_{\rm{o}}$ is large (0.15) and BBN constraints 
are discarded.  The exponential term in equation~(\ref{Eq:Gamma_th}) 
becomes significant because the $f_{\rm{G}}$ constraint forces 
$\Omega_{\rm{B}}$ to increase with $\Omega_{\rm{M}}$ allowing the 
product $\Omega_{\rm{M}}h$ to exceed 0.25.  This has been noted 
before (White \etal, 1996; Lineweaver \etal, 1997).  The SCDM model 
with $\Omega_{\rm{M}}$ = 1 and $h \approx 0.45$ is acceptable 
but the high value for the baryon-to-photon ratio, $\eta_{10} 
\approx 13$, is in conflict with the inferred primordial abundances 
of {\bf all} the light nuclides.  Note that, although the uncertainties 
are large, low values of $\eta_{10}$ are disfavored (see Figure 2).  

If we add the Dekel-Rees estimate of $\Omega_{\rm{M}}$ ($\Omega_{\rm{o}} 
\gsim 0.4$), the five-constraint fit favors somewhat higher values of 
$\Omega_{\rm{M}}$ and $\eta_{10}$ and slightly lower values of $h$.  In 
contrast, if instead we include the cluster estimate ($\Omega_{\rm{o}} = 
0.2 \pm 0.1$; Carlberg, 1997; cf. Carlberg, Yee, and Ellingson, 1997), we 
find a barely acceptable fit ($\chi^2_{\rm{min}}$ = 5.0 for 2 DOF, 92\% 
CL), which favors lower values of $\Omega_{\rm{M}}$ and $\eta_{10}$ and 
slightly higher values of $h$.

\subsection{Variations} 
 
Tilt in the primordial spectrum has been investigated in many papers 
(Liddle \etal, 1996a,b; White \etal, 1996; Kolatt and Dekel, 1997; White 
and Silk, 1996; Liddle and Viana, 1996).  We considered the effect of a 
moderate ``red tilt'' ($n = 0.8$ instead of $n = 1$).  This has the 
effect (see equation~\ref{Eq:Gamma_th}) of raising slightly the 68\% 
and 95\% contours in Figure 1.  With this tilt the $\Gamma$ constraint 
favors higher $\Omega_{\rm{M}}$, so that the SCDM model is allowed 
for $h$ up to nearly 0.5.  The favored likelihood range for $\eta_{10}$ 
is now also higher, though $\eta_{10} \approx 7$ is still allowed.   
However, the higher allowed range for $\eta$ does threaten the 
consistency of BBN.  Conversely, a ``blue'' tilt, $n > 1$ (Hancock 
\etal, 1994), would move the CR downward and allow models with 
$\Omega_{\rm{M}} \le 0.3$ at high $h$. 
 
Changing to a gas enhancement factor $\Upsilon = 1.3$ (modest 
positive enhancement of gas in clusters) instead of 0.9 does not 
change the contours in Figure 1 by much since $\Gamma$ is 
only weakly coupled to $\Omega_{\rm{B}}$ through the exponential 
term in $\Gamma$.  Although the effect is to lower the contours 
in Figure 1 slightly and to move downward the acceptable range 
for $\eta_{10}$, $\eta_{10} \le 4$ is still excluded, disfavoring 
the low D abundance inferred from some QSO absorbers and favoring 
a higher helium abundance than is revealed by the \hii-region data.  

The possibility that the fraction of cluster mass in baryons in 
galaxies, isolated stars, and machos $(f_{\rm{GAL}})$ might be larger 
-- even much larger -- than is implied by equation~(\ref{Eq:f_G/f_GAL}) 
would affect the CRs in much the same way as a {\it small} $\Upsilon$, 
favoring even higher values of $\Omega_{\rm{M}}$ and $\eta_{10}$. 

The $\Gamma$ constraint is crucial for our standard results favoring 
high $\Omega_{\rm{M}}$ and high $\eta$.  If, for example, we drop 
the $\Gamma$ constraint and in its place use the cluster estimate 
$\Omega_{\rm{o}} = 0.2 \pm 0.1$, low $\Omega_{\rm{M}}$ and low $\eta$ 
are now favored (see SHF).

The acceptability -- or not -- of the SCDM model depends crucially
on the choice of Hubble parameter.  SHF have experimented with 
replacing the standard constraint on H$_0$ with $h_{\rm{o}} = 
0.70 \pm 0.07$.  Now, the SCDM model is strongly excluded.

\subsection{$\Lambda$CDM MODELS} 
 
SHF have also considered models with nonzero $\Lambda$, limiting 
their investigation to the popular flat ($k$ = 0) ``$\Lambda$CDM'' 
models with $\Omega_\Lambda = 1 - \Omega_{\rm{M}}$, where 
$\Omega_\Lambda \equiv \Lambda /(3H_0^2)$.  For these models there 
are still only three free parameters and the four constraints discussed 
earlier are still in force, except that the product of the age and 
the Hubble parameter is a different function of $\Omega_{\rm{M}} = 
1 - \Omega_\Lambda$: $ t = 9.78\,h^{-1}f(\Omega_{\rm{M}};k=0)$ Gyr 
[Carroll, Press, and Turner, 1992, equation (17)].  For a given 
$\Omega_{\rm{M}} < 1$, the age is longer for the flat $(k = 0)$ 
model than for the $\Lambda = 0$ model.  The results differ very 
little from those in Figures 1 and 2.  The longer ages do allow the 
CRs to slide farther down toward large $h$ and small $\Omega_{\rm{M}}$.  
Because of the longer ages at low $\Omega_{\rm{M}}$ (high $\Omega_{\Lambda}$), 
$\Omega_{\rm{o}}$ from clusters can now be accepted as a fifth constraint.  
In this case (see SHF) large $\Omega_{\rm{M}}$ and small $h$ are now 
excluded while $\eta_{10} > 4$ is still favored strongly. 

\section{Conclusions} 
 
If BBN constraints on the baryon density are removed (or relaxed), 
the interaction among the shape-parameter $(\Gamma)$ constraint, 
the cluster baryon fraction ($f_{\rm{G}}$) constraint, and the 
value of $\eta_{10}$ assumes critical importance.  These constraints 
still permit a flat CDM model, but only as long as $h < 0.5$ is 
allowed by observations of $h$.  The $f_{\rm{G}}$ constraint means 
that large $\Omega_{\rm{M}}$ implies fairly large $\Omega_{\rm{B}}$.  
Therefore the exponential term in $\Gamma$ becomes important allowing 
$\Omega_{\rm{M}} = 1$ to satisfy the $\Gamma$ constraint.  However, 
values of $\eta_{10} \approx 8-15$ are required (see Figures 1 and 2).  
The best-fit SCDM model has $h \approx 0.45$ and $\eta_{10} \approx 
13$, which is grossly inconsistent with the predictions of BBN 
and the observed abundances of D, $^4$He, and $^7$Li.  For $h > 
0.5$ a fit to SCDM is no longer possible.  The SCDM model is 
severely challenged. 
 
The $\Gamma$ and age constraints also challenge low-density CDM 
models.  The $\Gamma$ constraint permits $\Omega_{\rm{M}} < 0.4$ 
only for high $h$, while the age constraint forbids high $h$, so 
$\Omega_{\rm{M}} \gsim 0.4$ is required.  The bound $\Omega_{\rm{M}} 
\gsim 0.4$ conflicts with the added cluster constraint $\Omega_{\rm{o}} 
= 0.2 \pm 0.1$ at the 98\% CL, suggesting strongly that there is 
additional mass not traced by light. 
 
Although a few plausible variations on the CDM models do not affect 
the constraints very much, removing the $\Gamma$ constraint would have 
a dramatic effect.  Both high and low values of $\Omega_{\rm{M}}$ would 
then be permitted.  The $\Gamma$ constraint plays a crucial role in our 
analysis.

At either low or high density, the situation remains about the same 
for the $\Lambda$CDM models.  Because the ages are longer, we can 
tolerate $\Omega_{\rm{M}} \approx 0.3$ for $h = 0.85$.  The $\Lambda$CDM 
model therefore accepts more easily the added constraint $\Omega_{\rm{o}} = 
0.2 \pm 0.1$.  Improved future constraints on $\Omega_\Lambda$ 
will come into play here. 
 
Having bounded the baryon density using data independent of 
constraints from BBN, we may explore the consequences for the 
light-element abundances.  In general, our fits favor large 
values of $\eta_{10}$ ($\gsim 6$) over small values ($\lsim 2$).  
While such large values of the baryon density are consistent with 
estimates from the Ly-$\alpha$ forest, they do create some tension 
for BBN.  For deuterium there is no problem, since for $\eta_{10} 
\gsim 6$ the BBN-predicted abundance, (D/H)$_{\rm{P}}$ $\lsim 3 \times 
10^{-5}$ (2$\sigma$), is entirely consistent with the low abundance 
inferred for some of the observed QSO absorbers (Tytler \etal, 1996; 
Burles and Tytler, 1996).  Similarly, the BBN-predicted lithium 
abundance, (Li/H)$_{\rm{P}} \gsim 2.5 \times 10^{-10}$ (2$\sigma$), 
is consistent with the observed surface lithium abundances in the 
old, metal-poor stars (allowing, perhaps, some minimal destruction 
or dilution of the prestellar lithium).  However, the real challenge 
comes from $^4$He where the BBN prediction for $\eta_{10} \gsim 6$, 
Y$_{\rm{P}} \gsim 0.248$ (2$\sigma$), is to be  contrasted with the 
\hii-region data which suggest Y$_{\rm{P}}$ $\lsim 0.238$ (OS, OSS).

\acknowledgements 

The research of G.S. at Ohio State is supported by DOE grant 
DE-FG02-91ER-40690.  N.H. is supported by the National Science 
Foundation Contract No. NSF PHY-9513835.  J.E.F. acknowledges 
fruitful visits to the Physics Department at Ohio State University 
and J.E.F. and G.S. acknowledge useful visits to the Aspen Center 
for Physics.


\begin{thebibliography}{}   

\bibitem[]{Bludman-97} 
Bludman, S. A.: 1997, {\it ApJ}, submitted, astro/ph 9706047. 
 
\bibitem[]{Bolte-Hogan-95}  
Bolte, M., and Hogan, C. J.: 1995, {\it Nature \bf 376}, 399.
 
\bibitem[]{Borgani-etal-97}  
Borgani, S., {\it et al.}: 1997, {\it New Astr. \bf 1}, 321.

\bibitem[]{Bureau-Mould-Staveley-Smith-96}  
Bureau, M., Mould, J. R., and Staveley-Smith, L.: 1996, {\it ApJ \bf 463}, 60. 
 
\bibitem[]{Burles-Tytler-96}  
Burles, S., and Tytler, D.: 1996, {\it ApJ \bf 460}, 584. 
 
\bibitem[]{Carlberg-97}  
Carlberg, R. G.: 1997, private communication. 
 
\bibitem[]{Carlberg-Yee-Ellingson-97}  
Carlberg, R. G., Yee, H. K. C., and Ellingson, E.: 1997, {\it ApJ \bf  478}, 462. 
 
\bibitem[]{Carroll-Press-Turner-92}  
Carroll, S. M., Press, W. H., and Turner, E. L.: 1992, {\it ARA\&A \bf  30}, 499. 
 
\bibitem[]{Carswell-etal-94}  
Carswell, R. F., Rauch, M., Weymann, R. J., Cooke, A. J., and Webb,  
J. K.: 1994, {\it MNRAS \bf 268}, L1. 
 
\bibitem[]{Cen-97}  
Cen, R.: 1997, {\it ApJ \bf 485}, 39. 
 
\bibitem[]{Chaboyer-Demarque-Kernan-Krauss-97}  
Chaboyer, B., Demarque, P., Kernan, P. J., and Krauss, L. M.: 1997, {\it ApJ}, 
submitted, astro-ph/9706128. 
 
\bibitem[]{Copi-Schramm-Turner-95}  
Copi, C. J., Schramm, D. N., and Turner, M. S.: 1995, {\it Phys. Rev. Lett.
\bf 75}, 3981. 
 
\bibitem[]{Cowan-McWilliam-Sneden-Burris-97}  
Cowan, J. J., McWilliam, A., Sneden, C., and Burris, D. L.: 1997, {\it ApJ \bf 480}, 246. 
 
\bibitem[]{DAntona-Caloi-Mazzitelli-97}  
D'Antona, F., Caloi, V., and Mazzitelli, I.: 1997, {\it ApJ \bf 477}, 519. 
 
\bibitem[]{Dekel-97}  
Dekel, A.: 1997, in {\it Galaxy Scaling Relations:  Origins, Evolution  
and Applications}, ed. L. da Costa (Berlin: Springer), in press, 
astro-ph/9705033. 
 
\bibitem[]{Dekel-Rees}  
Dekel, A., and Rees, M. J.: 1994, {\it ApJ \bf 422}, L1. 
 
\bibitem[]{Edmunds-94}  
Edmunds, M. G.: 1994, {\it MNRAS \bf 270}, L37. 
 
\bibitem[]{Evrard-97}  
Evrard, A. E.: 1997, {\it MNRAS \bf 292}, 289. 
 
\bibitem[]{Evrard-Metzler-Navarro-96}  
Evrard, A. E., Metzler, C. A., and Navarro, J. F.: 1996, {\it ApJ \bf 469}, 494. 
 
\bibitem[]{Fixsen-etal-96}  
Fixsen, D. J., Cheng, E. S., Gales, J. M., Mather, J. C., Shafer,  
R. A., and Wright, E. L.: 1996, {\it ApJ \bf 473}, 576. 
 
\bibitem[]{Hancock-etal-94}  
Hancock, S., Davies, R. D., Lasenby, A. N., Gutierrez de la Crux, C. M., 
Watson, R. A., Rebolo, R., and Beckman, J. E.: 1994, {\it Nature \bf 367}, 333. 
 
\bibitem[]{Hata-etal-95}  
Hata, N., Scherrer, R. J., Steigman, G., Thomas, D., Walker, T. P., 
Bludman, S., and Langacker, P.: 1995, {\it Phys. Rev. Lett. \bf 75}, 3977. 
 
\bibitem[]{Jimenez-97}  
Jimenez, R.: 1997,  
Invited lecture at the Cosmology School in Casablanca 97, 
astro-ph/9701222. 
 
\bibitem[]{Kolatt-Dekel-97}  
Kolatt, T., and Dekel, A.: 1997, {\it ApJ \bf 479}, 592.  
 
\bibitem[]{Kundic-etal-97}  
Kundi\'{c}, T., {\it et al.}: 1997, {\it ApJ \bf 482}, 75. 

\bibitem[]{Liddle-Lyth-Roberts-Viana-96}  
Liddle, A. R., Lyth, D. H., Roberts, D., and Viana, P. T. P.: 1996a, 
{\it MNRAS \bf 278}, 644. 
 
\bibitem[]{Liddle-Lyth-Viana-White-96}  
Liddle, A. R., Lyth, D. H., Viana, P. T. P., and White, M.: 1996b, 
{\it MNRAS \bf 282}, 281. 
 
\bibitem[]{Liddle-Viana-96}  
Liddle, A. R., and Viana, P. T. P.: 1996, in {\it Aspects of Dark Matter  
in Astro- and Particle Physics} (Heidelberg, September 1996), ed.  
H. V. Klapdor-Kleingrothaus and Y. Ramachers   
(Heidelberg: World Scientific), in press, astro-ph/9610215. 
 
\bibitem[]{Lineweaver-Barbosa-Blanchard-Bartlett-97} 
Lineweaver, C. H., Barbosa, D., Blanchard, A., and Bartlett, J. G.:  
1997, {\it A\&A \bf 322}, 365. 
 
\bibitem[]{Maddox-Efstathiou-Sutherland-96}  
Maddox, S. J., Efstathiou, G., and Sutherland, W. J.: 1996, {\it MNRAS  
\bf 283}, 1227. 
 
\bibitem[]{Nittler-Cowsik-97}  
Nittler, R. L., and Cowsik, R.: 1997, {\it Phys. Rev. Lett. \bf 78}, 175. 
 
\bibitem[]{Olive-Steigman-95}  
Olive, K. A., and Steigman, G.: 1995, {\it ApJS \bf 97}, 49. (OS) 
 
\bibitem[]{Olive-Skillman-Steigman-97}  
Olive, K. A., Skillman E. D., and Steigman, G.: 1997, {\it ApJ \bf 483}, 788. (OSS) 
 
\bibitem[]{Peacock-97}  
Peacock, J. A.: 1997, {\it MNRAS \bf 284}, 885. 
 
\bibitem[]{Peacock-Dodds-94}  
Peacock, J. A., and Dodds, S. J.: 1994, {\it MNRAS \bf 267}, 1020. 
 
\bibitem[]{Prantzos-96}  
Prantzos, N.: 1996, {\it A\&A \bf 310}, 106. 
 
\bibitem[]{Rugers-Hogan-96}  
Rugers, M., and Hogan, C. J.: 1996, {\it ApJ \bf 459}, L1. 
 
\bibitem[]{Songaila-Cowie-Hogan-Rugers-94}  
Songaila, A., Cowie, L. L., Hogan, C. J., and Rugers, M.: 1994, {\it Nature 
\bf 368}, 599. 

\bibitem[]{Steigman-Schramm-Gunn-77} Steigman, G., Schramm, D. N., 
and Gunn, J.: 1977, {\it Phys. Lett. \bf B66}, 202.

\bibitem[]{Steigman-Tosi-95}  
Steigman, G., and Tosi, M.: 1995, {\it ApJ \bf 453}, 173. 

\bibitem[]{Steigman-Hata-Felten-97}
Steigman, G., Hata, N., and Felten, J. E.:  1997, {\it ApJ}, submitted, 
astro-ph/9708016. (SHF)
 
\bibitem[]{Sugiyama-95}  
Sugiyama, N.: 1995, {\it ApJS \bf 100}, 281. 
 
\bibitem[]{Tammann-Federspiel-97}  
Tammann, G. A., and Federspiel, M. 1997,  
in {\it The Extragalactic Distance Scale}, ed. M. Livio, M. Donahue, and N. Panagia  
(Cambridge: Cambridge Univ. Press), 137. 
 
\bibitem[]{Tonry-Blakeslee-Ajnar-Dressler-97}  
Tonry, J. L., Blakeslee, J. P., Ajhar, E. A., and Dressler, A.:  
1997, {\it ApJ \bf 475}, 399. 
 
\bibitem[]{Tosi-96}  
Tosi, M.: 1996, in {\it From Stars to Galaxies:  
The Impact of Stellar Physics on Galaxy Evolution},  
ed. C. Leitherer, U. Fritze-von Alvensleben, and J. Huchra,  
ASP Conf. Series 98 (San Francisco: ASP), 299. 
 
\bibitem[]{Tytler-Fan-Burles-96}  
Tytler, D., Fan, X.-M., and Burles, S.: 1996, {\it Nature \bf 381}, 207. 
 
\bibitem[]{Walker-etal-91}  
Walker, T. P., Steigman, G., Schramm, D. N., Olive, K. A., and  
Kang, H.-S.: 1991, {\it ApJ \bf 376}, 51. 

\bibitem[]{Weinberg-72}  
Weinberg, S. 1972, {\it Gravitation and Cosmology} (New York: Wiley). 
 
\bibitem[]{White-Silk-96}  
White, M. and Silk, J. I.: 1996,  
{\it Phys. Rev. Lett. \bf 77}, 4704; erratum ibid., {\bf 78}, 3799. 
 
\bibitem[]{White-Viana-Liddle-Scott-96}  
White, M., Viana, P. T. P., Liddle, A. R., and Scott, D.: 1996, 
{\it MNRAS \bf 283}, 107. 
 
\bibitem[]{White-Navarro-Evrard-Frenk-93}  
White, S. D. M., Navarro, J. F., Evrard, A. E., and Frenk, C. S.:  
1993, {\it Nature \bf 366}, 429. 
  
\end{thebibliography}
\end{document}